# Vibrational zero point energy for H-doped Silicon

S. Zh. Karazhanov[1], M. Ganchenkova[2,3], and E. S. Marstein[1]

[1]Department for Solar Energy, Institute for Energy Technology, Instituttveien 18, 2027 Kjeller, Norway
[2]Department of Materials Science, National Research Nuclear University "MEPhI", 31 Kashirskoe sh, 115409, Moscow, Russia
[3]Department of Applied Physics, School of Science, Aalto University, PO Box 11000, FI-00076 Aalto, Espoo, Finland

**Abstract**
Most of the studies addressed to computations of hydrogen parameters in semiconductor systems, such as silicon, are performed at zero temperature T=0 K and do not account for contribution of vibrational zero point energy (ZPE). For light weight atoms such as hydrogen (H), however, magnitude of this parameter might be not negligible. This work is devoted to clarify the importance of accounting the zero-point vibrations when analyzing hydrogen behavior in silicon and its effect on silicon electronic properties. For this, we estimate the ZPE for different locations and charge states of H in Si. We show that the main contribution to the ZPE is coming from vibrations along the Si-H bonds whereas contributions from other Si atoms apart from the direct Si-H bonds play no role. It is demonstrated that accounting the ZPE reduces the hydrogen formation energy by ~0.17 eV meaning that neglecting ZPE at low temperatures one can underestimate hydrogen solubility by few orders of magnitude. In contrast, the effect of the ZPE on the ionization energy of H in Si is negligible. The results can have important implications for characterization of vibrational properties of Si by inelastic neutron scattering, as well as for theoretical estimations of H concentration in Si.

[1] Corresponding author: E-mail: smagulk@ife.no

## 1. Introduction

General procedure of theoretical determination of light atom thermodynamic parameters such as, e.g., the heat of formation and formation energy of defects, based on the density functional theory (DFT) correspond to zero temperature (T=0 K) and are therefore systematically underestimated. One reason for this is that these parameters derived from DFT computations do not include vibrational zero-point energy (ZPE) [1, 2]. ZPE for light weight elements in vacuum, as well as when doped or alloyed into other materials, might be few hundreds of milli-electron volts, which cannot be neglected [1-3]. A good estimate of the ZPE value is especially needed for correct estimation of the solution energy, and, consequently, the ionization energy, as well as the solubility of the light-weight elements in other materials. ZPE cannot be measured directly, because no molecule and/or atom can be at T=0 K. Consequently, the available experimentally measured values of ZPEs can be considered as average parameter estimated from a hybrid of experimental measurement and theoretical estimation.

In this work, we will focus on studies of H, which is one of the light weight elements. It is often present in many solids and plays key role in determining the electrical, optical, and structural properties of the materials [4, 5]. In Si, the well-studied elemental semiconductor, the role of H in determining its properties is manifold. For example, H can passivate deep level impurities [6], shallow donors [7-9] and shallow acceptors [10-12], interface states [13], can be a bistable defect in a complex with some group-V impurities [14], can enhance oxygen diffusion in $p$-Si [15], lead to formation of nanoclusters [16, 17], as well as a long list of other effects. Systematic theoretical studies of the ionization energy of H in Si have been performed (see, e.g., Refs. [18-23]). The local vibrational modes of different models of H in Si have also been studied by many authors (see, e.g., Refs. [23, 24]). However, despite the potential importance of ZPE in determining the H-related properties of Si, there is no systematic study of this point. In particular, the magnitude of the ZPE for atomic H in Si was not estimated for different size supercells for different charge states of H within different approximations. It is therefore not clear how it influences the solubility limit and ionization energy of H in Si, nor how different locations and charge states of H in Si, as well as the H-induced displacement of Si atoms influence the magnitude of the ZPE. The role of the Si atoms nearest to H in determining the ZPE is also not clarified. These issues would be important for determining the concentration and the ionization energies of H in Si as well as for the analysis of results obtained by inelastic neutron scattering (INS). Determination of the ZPE of H in Si is the topic of the present article.

## 2. Methods

The study has been performed by using the density functional theory (DFT) approach implemented in plane-wave-based code VASP [25]. The calculations used the generalized gradient approximation with Perdew–Burke–Ernzerhof exchange-correlation functional [26]. In order to describe the core electrons of the steel host atoms, the projector-augmented wave (PAW) method [33-35] has been employed. We have considered a unit cell of bulk Si of space group symmetry $F\bar{4}3m$. Using the lattice parameter $a$=5.392 Å of Ref. [27] as input, atomic positions in and volume of the unit cell have been relaxed and ground state parameter $a \approx 5.47$ Å has been found, which is in good agreement with the experimentally determined $a$ =5.43 Å [28]. The defect studies have been performed in 2×2×2, 3×3×3, 3×3×4, and 4×4×4 supercells, hereafter referred to according to the number of the atoms in the supercells as 64, 216, 288, and 512 atom supercells, respectively. It was needed to study convergence of the results. We have considered atomic H [Fig. 1 (a) and (b)] and molecular H$_2$ [Fig. 1 (c)]. Two locations of atomic H in the lattice of Si have been considered. One is at (or near to) the bond center (BC) [Fig. 1 (a)] and the other one is at the antibonding (AB) site [Fig. 1 (b)] in positively charged (H$^+$), neutral (H$^0$), and negatively charged (H$^-$) states. Location for the H$_2$ [Fig. 1(c)] has been selected based on previous studies (see, e.g., Refs. [18, 29]), which reported it as the most stable configuration compared to the lowest neutral atomic H.

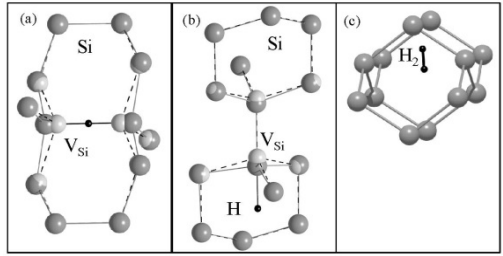

*Figure 1. Schematic presentation of H in Si at the (a) BC and (b) AB sites as well as (c) H$_2$ molecule. Circles marked by grey color correspond to Si atoms in the ideal lattice with H. Chemical bond between them is indicated by continuous grey color bond. White-grey color particles correspond to those Si atoms in the ideal lattice without H. Chemical bond between them is shown by dashes.*

In thermodynamic equilibrium between hydrogen atoms in a solid matrix and in gas atmosphere, the numbers of gas atoms occupying interstitial sites can be found by minimization of a relevant thermodynamic potential. Assuming constant temperature and pressure, the Gibbs free energy change, $\Delta G$, associated with the transfer of $N$ hydrogen atoms from the gaseous ambient environment into the silicon crystal can be expressed as

$$\Delta G(V,T) = \Delta E(V) + \Delta F_{vib}(V,T) + \Delta F_{el}(V,T) + PV - \mu_H N \text{ , (1)}$$

where $\Delta E(V)$ is the total energy change of the sample of volume $V$ at a given temperature $T$ and pressure $P$ due to hydrogen insertion into the lattice, while the last term describes the free energy decrease of the environmental gas in terms of the hydrogen chemical potential in the gas, $\mu_H$. $E(V)$ is the static energy, that depends on the unit cell volume and can be obtained from the full structural relaxation. $\Delta F_{el}$ is the contribution to the free energy from the electronic excitations, which can be neglected (Ref. [30]). $\Delta F_{vib}$ is the vibrational contribution the free energy:

$$\Delta F_{vib} = \frac{1}{2}\sum_{i=1}^{3}\hbar\omega_i + k_B T \sum_{i=1}^{3}\ln\left(1 - \exp\left(-\hbar\omega_i / k_B T\right)\right), \text{ (2)}$$

where $\hbar$ is the Planck´s constant and $\omega_i$ are the eigenfrequencies of hydrogen normal vibrations in the local potential well of the interstitial position. The first term in equation (2) describes the zero-point energy (ZPE) contribution, while the second one is the effect of thermal vibrations. For T=0 K it simplifies to

$$\Delta F_{vib} = ZPE = \frac{1}{2}\sum_{i=1}^{3}\hbar\omega_i \cdot \quad (3)$$

Since thermodynamic functions of Si are determined mostly by the vibrational degrees of freedom of the ions, knowledge of the vibrational spectra is very important and can be obtained from the first-principles calculations.

In order to estimate the contributions of the vibration free energy $\Delta F_{vib}$ (ZPE), the hydrogen eigenfrequencies were calculated from the Hessian matrices for the hydrogen potential energy surface in the equilibrium hydrogen positions using the finite differences approach implemented in VASP code. In eigenfrequency calculations, the lattice atoms were kept fixed in their relaxed positions.

The configuration energy $E^{con}[D]^q$ of defects in the charge state $q$ has been estimated by

$$E^{con}[D]^q = E_{tot}[D]^q - E_{tot}(Si). \quad (4)$$

$E_{tot}[D]^q$ and $E_{tot}(Si)$ are the total energy of the defective and ideal Si supercells. The solution energy $\Delta H^f[D]^q$ of defects in the charge state $q$ has been estimated by

$$\Delta H^f[D] = E^{con}[D]^q + \mu_{Si} - \mu_H + q(E_{VBM} + E_F). \quad (5)$$

Here $\mu_{Si}$ and $\mu_H$ are the chemical potentials for the elemental phases of Si and $H_2$. $\mu_H = 0.5\mu_{H_2}$ was taken as the H-rich limit. The calculated total energies of the Si crystal and $H_2$ molecules have been used as the respective chemical potentials. For H we got $\mu_H$ =3.36 eV. $E_{VBM}$ is the valence band maximum (VBM).

### 3. Results

Structural optimization of the lattice with defects has been performed for four different Monhkhorst-Pack [31] *k*-point samplings (2×2×2, 3×3×3, 4×4×4, and 5×5×5) showing that already the 4×4×4 mesh and 400 eV of cutoff energy ($E$cut) provides the accuracy of 10 meV for the H solution energy in all the considered configurations. This conclusion can be proved by the dependences presented in Figure 2. Figure 2 (a) presents the results obtained from studies related to the optimization of the lattice for the dependence of configuration energy $E^{con}$ calculated by Eq. (1) on $E$cut for the above-mentioned supercells containing one BC $H^0$ atom. Analysis shows that for $E$cut exceeding 400 eV, the variation of $E^{con}$ is less than 10 meV. So, $E$cut=400 eV has been considered as the optimal parameter and in further computations it was used as input. To find the optimal **k**-grid, $E^{con}$ has been calculated for different **k**-meshes [Fig. 2 (b)]. Analysis shows that for **k**-grids exceeding 4×4×4, the variation of $E^{con}$ is less than 10 meV. So, the 4×4×4 **k**-grid can be considered as the optimal one and was used as input in further computations for BC and AB $H^+$, $H^0$, and $H^-$ as well as for $H_2$ in Si.

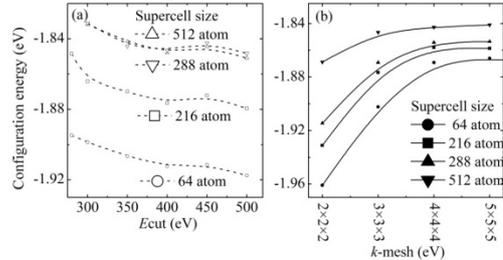

*Figure 2. Configuration energy for 64, 256, 288, and 512 atom Si supercells with a BC $H^0$ as a function of the (a) plane-wave-cut-off energy and **k**-mesh.*

In order to find the ground state and metastable states for different charge states and locations of atomic H in Si at different Fermi levels we have considered, first, 64 atom supercell containing H in BC and AB locations in the charge states 1+, 0 and 1-. The H configuration possessing smallest total energy has been considered as the ground state. The total energy analysis shows that $[E_{tot}(H_{BC})^0 - E_{tot}(H_{AB})^0]$ = -0.29 eV for neutral H and $[E_{tot}(H_{BC})^- - E_{tot}(H_{AB})^-]$ = +0.41 eV for negatively charged H. This result is in general agreement with previous theoretical studies (see, e.g., Ref. [32]) and shows that the ground state for $H^0$ is at the BC location whereas that for $H^-$ is at the AB location. Similar analysis has been performed for $H^+$ for which the ground state is at the BC site. Upon relaxation of the lattice, the $H^+$ atom, which initially was located at AB site, has been shifted to one of the nearest BC sites manifesting its instability for $H^+$.

The displacement of Si atoms caused by the incorporation of H has been studied [Fig. 1 (a) and (b)]. Analysis shows that the first and second nearest neighbors of H show the largest displacement from their equilibrium position corresponding to ideal lattice. Magnitude of the displacement is ~0.42 Å induced by BC $H^+$, ~0.45 Å for BC $H^0$, and ~ -0.31 Å for AB $H^-$, in agreement with previous findings (see, e.g., Ref. [18]). The rest Si atoms show small displacements, which are not clearly seen in Figs. 1(a) and (b). The displacement of Si atoms near to $H_2$ [Fig. 1(c)] was negligible. Magnitudes of the displacements of the host Si atoms located near to H are displayed in Fig. 3(a). Analysis shows that the displacements depend on both charge state and location of H. They increase with the variation of the charge state of H from negative to positive, neutral. Also, the Si atoms located near to BC $H^+$ and $H^0$ show a larger displacement than that located near to AB $H^-$. A large swing of the Si atoms around equilibrium position in the ideal lattice as a result of change of charge state of H from + to - causes both quantitative and qualitative variation of the volume. Namely, BC $H^0$ and $H^+$ causes increase of the unit cell volume whereas AB $H^-$ causes lattice contraction. These findings are in general agreement with earlier findings in other materials such as, e.g., $WO_3$ [33]. It should be noted that relaxation patterns for all the considered hydrogen atom locations do not change significantly when the supercell size changes from 64 to 512 atoms (Fig. 3(a)).

Upon changing the environment and location of H in a lattice, its vibrational spectra and ZPE will be changed. Vibrational spectra for the $H_2$ molecule in vacuum [Fig. 3(b)] have been calculated and a peak at 4271 cm$^{-1}$ has been found, which differs from the experimentally determined value of 4161 [34] cm$^{-1}$ only by ~2.6 %. The incorporation of the hydrogen molecule into silicon matrix leads to its vibrational peak shift to 3649 cm$^{-1}$ [Fig. 3(b)] manifesting $H_2$ interaction with surrounding silicon environment. The calculated peak value, even though somewhat larger than, but still is in a reasonable agreement with the experimentally measured 3601 cm$^{-1}$ [35] at T=295 K and 3618 [35] at T=10 K showing only 1.3 % and 0.9% difference, respectively. The calculated ZPE for $H_2$ in vacuum, 0.274 eV, corresponds well to the experimentally measured 0.270 eV [1]. Therefore, the calculated ZPE for molecular $H_2$ in Si 0.314 eV is expected to be quite accurate.

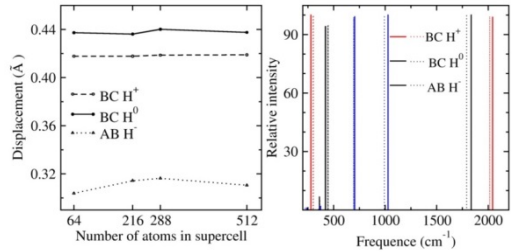

*Fig. 3. (a) Displacement of Si atoms located nearest to BC $H^+$, $H^0$, and AB $H^-$ as a function of the supercell size. For AB H- the displacement is negative and its absolute value is presented. (b) Vibrational spectra for BC $H^+$, $H^0$, and AB $H^-$ in Si obtained from ab initio calculations for 64 (solid lines) and 216 (dashes) atom supercells.*

Further, we have studied the vibrational spectra for BC $H^+$, $H^0$, and AB $H^-$ [Fig. 3(b)]. Analysis shows that the calculated vibrational spectra for these H configurations are in good agreement with experimental data and previous theoretical results. For BC $H^+$ we got two well-defined peaks at 2043 and 275 cm$^{-1}$, which are in line with experimentally determined 1998 cm$^{-1}$ (Ref. [36]) and previous theoretical estimates, which yield 1994 cm$^{-1}$ and a wag mode at 309 cm$^{-1}$ (Ref. [22]) by using the software SIESTA, 2203 cm$^{-1}$ (Ref. [37]) within DFT by Gaussian basis functions and 2210 cm$^{-1}$ (Ref. [18]) within the local-density approximation (LDA). The calculated peaks of large intensity for BC H$^0$ are located at 422 cm$^{-1}$ and 1839 cm$^{-1}$. The high frequency peak agrees well with calculated 1780 cm$^{-1}$ and 534 cm$^{-1}$ (Ref. [37]), with 1813 cm$^{-1}$ (Ref. [24]) and 1945 cm$^{-1}$ (Ref. [18]). Some peaks of small intensity are seen at smaller frequencies at 365 cm$^{-1}$ and 157 cm$^{-1}$. Vibrational peaks of high intensities for AB H$^-$ are located at 703 cm$^{-1}$ and 1030 cm$^{-1}$.

The ZPE values for different charge states of atomic H in BC and AB locations have been calculated for different size supercells [Table I]. Analysis shows that the difference in the ZPE corresponding to these supercells is less than 5 meV, which means that the calculated ZPE values are not sensitive to the supercell size starting from 64 atom size. However the charge state and the precise location of the atomic H in the lattice influence on magnitude of ZPE. For neutral BC H$^0$, the calculated ZPE is 162.60 meV, which is somehow smaller than 173.85 meV for BC H$^+$ and larger than 151.01 meV for AB H$^-$.

Table I. ZPE (meV) for the 64, 216, 288, and 512 atom supercells of Si with a BC H$^+$, H$^0$, and AB H$^-$. In the last two columns are ZPE for Si atoms in ideal lattice, which in defective lattice are located near to H.

|     | BC H$^0$ | | BC H$^+$ | | AB H$^0$ | | AB H$^-$ | | H$_2$ | Ideal lattice |
|---|---|---|---|---|---|---|---|---|---|---|
|     | H | Si | H | Si | H | Si | H | Si | H$_1$ | Si |
| 64  | 163 | 61 | 174 | 57 | 148 | 54 | 151 | 61 | 314 | 64 |
| 216 | 166 | 62 | 171 | 58 | 149 | 55 | 147 | 60 |     | 65 |
| 288 | 167 | 62 |     |    |     |    |     |    |     | 64 |

Our analysis of the vibrational contribution of the Si atoms surrounding H to the change of the free energy after the hydrogen atom incorporation into the silicon matrix shows that it is negligibly small. Indeed, as can be seen in Table I, the difference of ZPE's of the Si atoms in the cells with and without H is less than 10 meV and the major contribution to the magnitude of the $\Delta F_{vib}$ is coming from Si-H vibrations [see video, in supplementary information]. This result has important implications in the study of the vibrational properties of H in Si by INS. Namely, it shows that INS can provide reliable information about the locations of atomic H in different charge states, since one can neglect with the local atomic environment vibration contribution to the change of the vibrational spectrum due to hydrogen presence.

Figure 4 displays the solution energy for BC H$^+$ and H$^0$ as well as AB H$^-$ in Si calculated for two cases, one where the ZPE is taken into consideration, and another where the ZPE is neglected. The analysis shows that the ZPE results in a decrease of formation energy for H$^+$, H$^0$, and H$^-$ by ~0.1, 0.11, and 0.13 eV, respectively. However, the energy levels of the defect E(+/0) and E(0/-) have only been changed by less than ~50 meV leading us to conclusion that accounting zero-point vibration energy does not provide noticeable difference for electronic properties description of H in silicon. Since the concentration ($S$) of BC H$^+$ and AB H$^-$ depends on the $\Delta H_f(D)$ (Eq. (5)), the magnitude of the ZPE is expected to influence on [$S$]:

$$[S] \propto \exp\left[-\frac{\Delta H_f(D)}{k_B T}\right] \qquad (4)$$

Here $k_B$ is the Boltzmann constant and $T$ is the temperature. Since $\Delta H_f(D)$ depends on the position of the Fermi level $E_F$ [Eq. (5)], one can say that [$S$] depends on both the Fermi level and the ZPE. We have studied the relation of [$S$] in two cases, again either including or neglecting the effect of the ZPE [Fig. 4(b)]. The analysis shows that at moderate temperatures 300-500 K, the underestimation can be between one and two orders of magnitude.

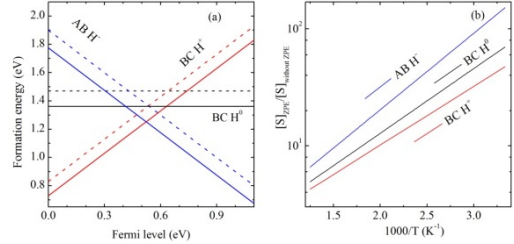

Fig. 4. (a) Formation energy calculated by neglecting (– – –) and including (—) ZPE. (b) Relation of concentration of BC H$^+$, H$^0$, and AB H$^-$ calculated by including ZPE into consideration to that neglecting ZPE.

### 4. Conclusion

In summary, we have studied vibrational zero point energy (ZPE) of H-doped Si by first-principles calculations. We have established the locations for ground and metastable states of H in Si in different charge states. We estimated the magnitude of the ZPE, which is in the range 0.15-0.17 eV, and it depends on the location and charge state of H in Si. The main contribution to the ZPE is coming from vibrations along the Si-H bonds. We show that the impact of other contributions from Si atoms located near to H to the ZPE is negligible. The ZPE noticeably influences the solution energy and solubility limit of H in Si at temperatures 300-500K, but it changes its ionization energies only moderately.

**Acknowledgments**. This work has received financial and supercomputing support from the Research Council of Norway through the FME project.

**Supporting Information.** A video showing vibration of atomic H in bond center and antibonding locations.